\documentclass[epj,final]{svjour}
\begin{document}
\title{A linear potential in a light cone QCD inspired model}
\author{Hans-Christian Pauli}
\institute{Max-Planck-Institut f\"ur Kernphysik, D-69029 Heidelberg }
\date{20 December 2003}  
\abstract{%
   The general equation from previous work is specialized
   to a linear potential $V(r)=-a+F\,r$ acting in the space of
   spherically symmetric S wave functions. 
   The fine and hyperfine interaction creates then a 
   $\frac1r$-dependence in the effective potential energy 
   equation and a position dependent mass $\widetilde m(r)$ 
   in the effective kinetic energy of the associated 
   Schr\"odinger equation.
   The results are compared with the available experimental 
   and theoretical spectral data on the $\pi$ and $\rho$. 
   Solving the eigenvalue problem within the analytically tractable 
   Airy-function approach induces a certain amount of 
   arbitrariness (fudge factors). 
   Despite of this, the agreement with experimental data is good
   and partially better than other calculations,
   including Godfrey and Isgur \cite{GodIsg85} 
   and Baldicchi and Prosperi \cite{BalPro02}.
   The short comings of the present model can be removed easily 
   in more elaborate work. 
} 
\maketitle
\section{The S-state Hamiltonian}
For spherically symmetric S states the previously derived Hamiltonian
reduces in Fourier approximation to \cite{Pau03a,Pau03b,Pau03c}
\begin{eqnarray}
   \begin{array}{lcl}
   H &=& \phantom{-}
   \frac{\vec p^2}{2m_r} + V + V_{hf}+ V_{K}+ V_{D}  
\,,\\ 
   V_\mathrm{hf} &=& \phantom{-}
   \frac{\vec\sigma_1\vec\sigma_2}{6m_1 m_2} \nabla^2 V   
\,,\\  
   V_\mathrm{K} &=& \phantom{-} 
   \frac{V}{m_1 m_2}  \vec p^2 
\,,\\  
   V_\mathrm{D} &=& -\Big[\frac{V}{16m_1 m_2} \vec p^2 +
   \frac{\nabla^2V}{4m_1 m_2}\Big] \big(\frac{m_1}{m_2}+\frac{m_2}{m_1}\big)  
\,.\end{array}
\label{eq:1}\end{eqnarray}
There are no more interactions than the central potential, 
the hyperfine, the kinetic, and the Darwin interaction, 
but also no less.
For s-states the total spin squared is a good quantum number
$\vec S^2=[(\vec\sigma_1+\vec\sigma_2)/2]^2=S(S+1)$, thus
\begin{eqnarray}
   \vec\sigma_1\vec\sigma_2&=& 2S(S+1) - 3 =
   \left\{\begin{array}{rll}
     +1, & \mbox{ for } S=1, &\mbox{ triplet}, \\
     -3, & \mbox{ for } S=0, &\mbox{ singlet}.\end{array} \right. 
\label{eq:2}\end{eqnarray}
Because it is shorter, $\vec\sigma_1\vec\sigma_2$ is kept 
explicit in the equations as an abbreviation for Eq.(\ref{eq:2}).
Choosing a linear potential,
\begin{eqnarray}
   V(r) &=& - a + F\,r 
\,,\end{eqnarray}

with the force parameter $F$, often called string tension $\sigma$,
the Hamiltonian (\ref{eq:1}) becomes a non-local Schr\"odinger equation
with a $\frac1r$-potential
\begin{eqnarray*}
   \begin{array}{rcl@{}c@{} l ll ll ll}
   H &=&\big[& & \frac{1}{2m_r} &+& \frac{V(r)}{m_1m_2} 
     &-& \frac{V(r)}{16m_1m_2}\big(\frac{m_1}{m_2}+\frac{m_2}{m_1}\big)
     &\big]& \vec p ^2
   \\
     &+&\big[& & & & Fr &-&  a &\big]& 
   \\
     &+&\big[& & & & \frac{F}{3m_1m_2}\vec\sigma_1\vec\sigma_2 
                 &-& \frac{F}{2m_1m_2}\big(\frac{m_1}{m_2}+\frac{m_2}{m_1}\big) 
     &\big]& \frac{1}{r}
\,,\end{array}
\end{eqnarray*}
since 
\begin{eqnarray}
   \nabla^2 V(r) &=& \frac1r\frac{d^2}{dr^2} rV(r) = 2\frac{F}{r} 
\,.\end{eqnarray}
Shaping notation, the Hamiltonian is written as
\begin{eqnarray}
   H &=& \phantom{-}
   \frac{\vec p^2}{2\widetilde m_r(r)} + Fr - a 
\nonumber\\ && -
   \frac{\beta}{r}\left(\frac{1}{2}\left[\frac{m_1}{m_2}+
   \frac{m_2}{m_1}\right]-\frac{\vec\sigma_1\vec\sigma_2}{3}\right)  
\,.\label{eq:5}\end{eqnarray}
The dimensionless `coupling constant' is
\begin{eqnarray}
   \beta =\frac{F}{m_2 m_1}
\,.\end{eqnarray}
The spin-averaged potential energy for equal masses,
\begin{eqnarray}
   V_\mathrm{av}\equiv \frac{3V_\mathrm{t}+V_\mathrm{s}}{4} =
   Fr - a - \frac{\beta}{r}
\,,\end{eqnarray}
has an attractive Coulomb potential. 
It has its origin in the Darwin term.
The non locality of the Hamiltonian resides in the position
dependent mass
\begin{eqnarray}
   \frac{m_r}{\widetilde m_r(r)} &=&
   1 + \frac{V(r)}{8(m_1+m_2)}
   \Big[16-\frac{m_1}{m_2}-\frac{m_2}{m_1}\Big]
.\end{eqnarray}
To solve this Hamiltonian, one must go on a computer.

The Hamiltonian in Eq.(\ref{eq:5}) 
looks like a conventional instant form Hamiltonian as obtained 
by quantizing the system at equal usual time.
But it must be emphasized that it continues to be 
a genuine front form or light cone Hamiltonian \cite{BroPauPin98},
derived from the latter by a series of exact unitary 
transformations \cite{Pau03a,Pau03b}.

\section{The model Hamiltonian and its parameters}
In this first round, I try to avoid to go on the computer as far
as possible, by the following reason.
The parameters in the theory must be determined from experiment,
and this turns out as a non trivial, strongly non linear problem.
In order to get a first and rough estimate, 
the Hamiltonian is simplified here
until it has a form which is amenable to analytical solution.
Therefore, all in-tractable terms in the above will be replaced 
here by mean values and related to the experimentally accessible 
mean square radius $\langle r^2\rangle$ \cite{PovHue90}. 

\begin{table} [t]
   \caption{\label{tab:model}
   Model parameters in GeV. Note: $[f^*]=1$.
}\begin{tabular}{||@{\ }c@{\,}c@{\ }||c@{\ }|c@{\ }|c@{\ }|c@{\ }||c|c||}
 \hline\hline 
 $f^*_r$ & $f^*_i$ & $m_{d,u}$ & $m_s$ & $m_c$  & $m_b$  &  $a$   &  $10F$  \\ 
 \hline\hline 
  2 & 1 & 0.4259 & 0.5553 & 1.8152 & 5.2505 & 1.1317 & 2.1454 \\
 \hline\hline
\end{tabular}
\end{table}
In effect, the substitution 
\begin{eqnarray}
   \widetilde m_r(r)\Longrightarrow \widetilde m_r
\,,\label{ieq:8}\end{eqnarray}
is the only true assumption in the present model.
I consider thus the model Hamiltonian, 
\begin{eqnarray}
   H &=& \phantom{-}
   \frac{\vec p^2}{2\widetilde m_r} + Fr -\widetilde a + 
   \widetilde c\vec\sigma_1\vec\sigma_2 
\,,\label{eq:9}\end{eqnarray}
with the abbreviations 
\begin{eqnarray} \lefteqn{\hspace{-27ex}
   \widetilde c \hspace{2ex} = \frac{\beta}{3} \Big\langle \frac 1 r\Big\rangle 
   \hspace{6ex}=
   \frac{F}{3m_1m_2}\Big\langle \frac 1 r\Big\rangle 
\,,}\\ \lefteqn{\hspace{-27ex}
   \widetilde a \hspace{2ex} = a + \Big\langle \frac 1 r\Big\rangle 
   \frac{\beta}{3}
   \hspace{2ex} = a + \frac{3\widetilde c }{2}
   \Big(\frac{m_1}{m_2}+\frac{m_2}{m_1}\Big) 
\,,}\label{eq:16}\\ \lefteqn{\hspace{-27ex}
   \frac{m_r}{\widetilde m_r} =
   1 + \frac{\phantom{8}(F\langle r\rangle -a)}{8(m_1+m_2)}
   \Big[16-\frac{m_1}{m_2}-\frac{m_2}{m_1}\Big]
\,.}\end{eqnarray}
Its eigenvalues are 
\begin{eqnarray}
   E_n &=& -\widetilde a + \omega\,\xi_0 + \omega\,\eta_n +
   \widetilde c\vec\sigma_1\vec\sigma_2
   \,,\quad 
   \omega = \bigg[\frac{F^2}{2\widetilde m_r}\bigg]^{\frac13} 
,\end{eqnarray}
with the negative zeros of the Airy functions  
$\xi_n$ and $\eta_n = \xi_n-\xi_{0}$. A few ones are tabulated
in App.~\ref{asec:1}.
The invariant mass squares 
\begin{eqnarray}
   M_n^2 &=& \big(m_1+m_2\big)^2  
\nonumber\\ &+&
   2\big(m_1+m_2\big)
   \big(-\widetilde a + \xi_0\omega + \eta_n \omega +
   \widetilde c\vec\sigma_1\vec\sigma_2 \big)
\,,\end{eqnarray}
are then related to experiment.

For equal masses $m_1=m_2=m$, the model has the 3 parameters 
$m$, $F$ and $a$.
One thus needs 3 empirical data to determine them. I choose:
\begin{eqnarray}
   \begin{array}{rcl}
   M^2_{d \bar u,t1} &=& 4m^2 
   + 4m\big(-\widetilde a+\xi_0\omega + \phantom{3}\widetilde c
   + \eta_1\omega\big)
\,,\\     
   M^2_{d \bar u,t0} &=& 4m^2 
   + 4m\big(-\widetilde a+\xi_0\omega + \phantom{3}\widetilde c\big)
\,,\\     
   M^2_{d \bar u,s0} &=& 4m^2 
   + 4m\big(-\widetilde a+\xi_0\omega - 3\widetilde c\big)
\,.\end{array}
\label{eq:13}\end{eqnarray}
The spectrum is labeled self explanatory by the flavor composition 
$M_n = M_{d\bar u,tn}$ or $M_n = M_{d\bar u,sn}$, 
for singlets or triplets, respectively. 
The triple chosen in Eq.(\ref{eq:13}) exposes a certain asymmetry.
The excited $\rho$ is chosen since its experimental limit 
of error is very much smaller than the one for the corresponding 
$\pi$ state. 
Only its ground state mass  
is known very accurately, \textit{i.e.}
$m_{\pi^+}=139.57018\pm0.00035\mbox{ MeV}$. 
In the present work only the first 4 digits are used.
For equal masses, the above abbreviations become
\begin{eqnarray} 
   \begin{array} {lcl}
   \widetilde c &=&  
   \frac{F}{3m^2}\Big\langle \frac 1 r\Big\rangle 
   \,, \\  
   \widetilde a &=& a + 3\widetilde c
   \,,\\  
   \frac{m_r}{\widetilde m_r} &=& 
   1 + \frac{7(F\langle r\rangle -a)}{8m}
   \,.\end{array}
\end{eqnarray}
The experiment defines 2 certainly positive differences:
\begin{eqnarray}
   \begin{array}{rclcl}
   X^2 &=& \phantom{\frac{\xi_0}{\eta_1}}
   M^2_{d \bar u,t1} - M^2_{d \bar u,t0} &=& 4m
   \phantom{[}\eta_1\omega
\,,\\     
   Y^2 &=& \phantom{\frac{\xi_0}{\eta_1}}
   M^2_{d \bar u,t0} - M^2_{d \bar u,s0} &=& 4m
   \phantom{[}4\widetilde c 
   \,.\end{array}
\label{eq:21}\end{eqnarray}
A third one will be constructed by the observation that
$\frac{\xi_0}{\eta_1}X^2-\frac32 Y^2-M^2_{d \bar u,s0}=4m a - 4m^2$. 
Keeping in mind that 
$\omega ^3= F^2/\widetilde m$,
one can remove trivial kinematic factors and define
3 experimental quantities $B$, $C$ and $D$ by 
\begin{eqnarray} 
   \begin{array}{lcl}
   B^2 &=& \frac14\big(\frac{\xi_0}{\eta_1}X^2 - 
   \frac32 Y^2 - M^2_{d \bar u,s0}\big) = ma-m^2
   \,,\\     
   C &=& 
   \frac{3}{16\langle\frac{1}{r}\rangle} Y^2= \frac{F}{m}
   \,,\\ 
   D^4 &=& \frac{2}{3Y^2\eta_1^3}\langle\frac{1}{r}\rangle
   \big(X^2\big)^3 = 
   8m^2 + 7mF\langle r\rangle  - 7ma 
   \,.\end{array}
\label{eq:18}\end{eqnarray}
Substituting $F=mC$ and $ma=B^2+m^2$ gives 
\begin{eqnarray*} 
   D^4 &=& [1+7\langle r\rangle C]\,\big(m^2\big)^2 -7B^2\,m^2  
\,,\end{eqnarray*}
a quadratic equation with the solution 
\begin{eqnarray}  
   m^2 &=& 
   \frac{7B^2}{2[1+7\langle r\rangle C]}
   \Bigg[1 + 
   \sqrt{1+\frac{4D^4}{49B^4}[1+7\langle r\rangle C]} \Bigg]
.\end{eqnarray}
Having $m$, the $F$ and $a$ are then calculated from (\ref{eq:18}). 

\begin{table}[t]
   \caption{\label{tab:fudge}
   Dependence on the fudge factors.
}\begin{tabular}{||c|c||@{\ }c@{\ }|@{\ }c@{\ }|@{\ }c@{\ }|@{\ }c@{\ }||c|c||}
\hline\hline
 $f^*_r$&$f^*_i$&$m_{d,u}$&$m/\widetilde m $&  $a$  &  10$F$   & 
 4$^1\mathrm{S}_0$ & 4$^3\mathrm{S}_1$\\ 
\hline\hline 
  1 &  4 &  1.891 &  0.216 &  2.050 &  1.191 & 1.9936 & 2.1320 \\ 
  4 &  4 &  0.550 &  1.894 &  1.097 &  1.385 & 1.9936 & 2.1320 \\ 
\hline 
  1 &  1 &  0.816 &  0.390 &  1.185 &  2.056 & 1.9936 & 2.1320 \\ 
  2 &  1 &  0.426 &  1.315 &  1.132 &  2.145 & 1.9936 & 2.1320 \\ 
  4 &  1 &  0.215 &  5.018 &  1.611 &  2.171 & 1.9936 & 2.1320 \\ 
 10 &  1 &  0.086 &  30.95 &  3.563 &  2.178 & 1.9936 & 2.1320 \\ 
 20 &  1 &  0.043 &  123.5 &  6.992 &  2.179 & 1.9936 & 2.1320 \\ 
 40 &  1 &  0.022 &  493.9 & 13.918 &  2.179 & 1.9936 & 2.1320 \\ 
\hline\hline
\end{tabular}
\end{table}
With Airy functions, moments of different powers in $r$ are somewhat difficult
to evaluate. Therefore Gaussian weights are used, which give
\begin{eqnarray}  
   \frac{\langle r \rangle^2}{\langle r^2 \rangle} &=& 
   \frac{8}{3\pi} 
\,,\qquad  
   \Big\langle\frac 1 r \Big\rangle^2 \langle r^2 \rangle = \frac{6}{\pi}  
\,.\end{eqnarray}
In order to allow for corrections due to the true wave functions
I introduce two fudge factors $f^*$ according to 
\begin{eqnarray}  
   \begin{array} {rcl} 
   \langle r \rangle  &=& 
   \sqrt{\frac{8\langle r^2 \rangle_\pi}{3\pi}} f^*_r
   \,,\\
   \big\langle\frac 1 r \big\rangle &=& 
   \frac{4}{\pi \langle r \rangle}  f^*_i
   \,.\end{array} 
\end{eqnarray}
Since all masons have about the same size \cite{PovHue90}, 
by order of magnitude, these numbers are kept universal.
The fudge factors are introduced here 
to account, in some global fashion, for the tremendous simplification 
introduced by replacing Eq.(\ref{eq:5}) with (\ref{eq:9}).
Some large scale variations of $f^*_r$ and $f^*_i$ are compiled
in Table~\ref{tab:fudge}. 
The mass spectra including the ground states vary very little with
the fudge factors. Any variations would show up the fastest for 
the high excitations. 
For this reason, the masses for $n=4$ are included in the table.
I do not understand this insensitivity from a mathematical 
or numerical point of view. 
The major effect of $f^*_r$ is
the ease by which one can change the quark mass.
A value of $f^*_r\sim40$ leads to the 20~MeV for the quark mass 
quoted in \cite{BalPro02}.
In the present model, the values $f^*_r=2$ and $f^*_i=1$ are taken
without seeking an optimum.

In principle, one could determine the heavier quark masses analytically
from the hyperfine splittings.
The so obtained results are, however, not very reasonable, since
experimental numbers are not sufficiently accurate. 
Therefore, I determine them numerically from  
$M_{u\bar s,s0}$, $M_{u\bar c,s0}$ and $M_{u\bar c,s0}$ and compile 
them in Table~\ref{tab:model}. 
The force parameter $F\sim 1100 \mbox{ MeV/fm}$ 
is in line with currently used string tensions. 

\section{Results and Discussion} 
\begin{table}[t]
   \caption{\label{tab:ud}S wave spectra in GeV for light unflavored mesons. 
}\begin{tabular}{||r|l|l||l|l|l||}
 \hline\hline
 n & $^1\mathrm{S}_0$ Singlets& $\pi^+ (u\bar d)$  &n
   & $^3\mathrm{S}_1$ Triplets& $\rho^+(u\bar d)$    \\
   & Experiment$^1$ & Theory    &   & Experiment$^1$& Theory    \\
 \hline\hline
 1 & 0.1396(0)      & 0.1396    & 1 & 0.7685(6)     & 0.7685    \\ 
   &                & 0.1396$^2$&   &               & 0.7711$^2$\\
   &                & 0.150$^3$ &   &               & 0.769$^3$ \\
   &                & 0.497$^4$ &   &               & 0.846$^4$ \\
 2 & 1.300(100)     & 1.2550    & 2 & 1.465(25)     & 1.4650    \\ 
   &                & 1.2650$^2$&   &               & 1.4650$^2$\\
   &                & 1.300$^3$ &   &               & 0.769$^3$ \\
   &                & 1.326$^4$ &   &               & 1.461$^4$ \\
 3 & 1.795(10)      & 1.6878    & 3 & 1.700(20)$^a$ & 1.8493    \\ 
   &                & 1.7950$^2$&   &               & 1.9230$^2$\\
   &                & 1.880$^3$ &   &               & 2.000$^3$ \\
   &                & 1.815$^4$ &   &               & 1.916$^4$ \\
 4 & ----           & 1.9936    & 4 & 2.150(17)     & 2.1320    \\ 
   &                & 2.1620$^2$&   &               & 2.2912$^2$\\
 \hline\hline
 \multicolumn{6}{l}{ %
   $^1$Hagiwara  \textit{etal} \cite{RPP02},\ %
   $^2$Zhou and Pauli \cite{ZhoPau03b}.\ %
}\\
 \multicolumn{6}{l}{ %
   $^3$Godfrey and Isgur \cite{GodIsg85},\ %
   $^4$Baldicchi and Prosperi \cite{BalPro02} (a),\ %
}\\
 \multicolumn{6}{l}{ %
   $^a$Could be a D state \cite{AniAniSar00}.
}\\
\end{tabular}
\end{table}
\begin{table}[t]
\caption{\label{table:us} 
   S wave spectra in GeV for strange mesons.}
\begin{tabular}{||l|l@{\,}|l||l|l|l||}
 \hline\hline
 n & Experiment$^1$ & Theory    & n & Experiment$^1$& Theory     \\
 \hline 
 \multicolumn{3}{||c@{\,}||}{$^1\mathrm{S}_0$ Singlets $K^+(u\bar s$)}   &
 \multicolumn{3}{  c||}{$^3\mathrm{S}_1$ Triplets $K^{*+}(u\bar s$)}\\
 \hline\hline
 1   & 0.493677(16) & 0.4937    & 1 & 0.89166(26)   & 0.8651    \\ 
     &              & 0.6048$^2$&   &               & 0.8917$^2$\\
     &              & 0.47$^3$  &   &               & 0.90$^3$  \\
 2   & 1.460$^a$    & 1.3943    & 2 & 1.629(27)$^b$ & 1.5649    \\ 
     &              & 1.5480$^2$&   &               & 1.6808$^2$\\
     &              & 1.45$^3$  &   &               & 1.58$^3$  \\
 3   & 1.830$^a$    & 1.8266    & 3 & ---           & 1.9598    \\ 
     &              & 2.1040$^2$&   &               & 2.6242$^2$\\
     &              & 2.02$^3$  &   &               & 2.11$^3$  \\
 4   & ----         & 2.1370    & 4 & ---           & 2.2520    \\ 
 \hline\hline
 \multicolumn{6}{l}{ %
   $^1$Hagiwara  \textit{etal} \cite{RPP02},\ %
   $^2$Zhou and Pauli \cite{ZhoPau03b}.\ %
}\\
 \multicolumn{6}{l}{ %
   $^3$Godfrey and Isgur \cite{GodIsg85},\ %
}\\
 \multicolumn{6}{l}{ %
   $^a$To be confirmed; $^b$$J^P$ not confirmed.
}\\
\end{tabular}
\end{table}

\textbf{Unflavored light mesons}.
The results for the $\pi$--$\rho$ system 
are compiled in Table~\ref{tab:ud}.
The experimental points are taken from from 
Hagiwara \textit{et al}~\cite{RPP02}. 
It is no surprise that theory and experiment coincide for
the $\pi^+$, the $\rho^+$ and the $\rho^+(1450)$, because
these data have been used to determine the parameters.
More surprizing is, that the model reproduces 
the huge mass of the excited pion within the error limit. 
This solves for me a long standing puzzle:
Why is it that the ground state has a mass of 140~MeV,
while the first excitation with its 1300~MeV
is different by an order of magnitude?~---  
The answer is the usual one:
Two large scales interfer destructively
for the ground and constructively for the excited
state. The two scales are the depth $a\sim 1100\mbox{ MeV}$
and the string constant  $F\sim 1100 \mbox{ MeV/fm}$.

The remaining three experimental masses of the $\pi$-$\rho$ sector
agree reasonably well with the calculation.
There is no confirmed datum for the second
excited $\rho^+$ ($3^3$S$_1$). 
The third excited $\rho^+$ ($4^3$S$_1$) deviates from experiment
by 18 MeV, which is almost wihin the limits of error.
The second excited $\pi^+$ ($3^1$S$_0$) deviates from experiment
by 107 MeV.
The table includes also a comparison with other theoretical calculations.
It includes the results from a recent  
oscillator model \cite{ZhoPau03b}. 
Their model is even simpler than the present one: 
it works with a hyperfine splitting, only, 
but suppresses the mechanism of a position dependent mass.
Despite this, their results differ little from the present ones.
I have included also the results from the pioneering work of 
Godfrey and Isgur \cite{GodIsg85} as a prototype of a phenomenological model, 
and from a recent advanced calculation by 
Baldicchi and Prosperi \cite{BalPro02}. 
Note that either of these models have not much 
in common with the present one. 
The present model gives a good description for the pion, 
particularly the small mass of the physical pion is reproduced.
All other potential models, including even Baldicchi and Prosperi,
have the wellknown difficulties with that.
I would have loved to compare also with Lattice Gauge Calculations
but of course no data are available for exited states, particularly
not for such light systems as the pion and the rho.
Note that the effort on a pocket calculator is ridicuously small 
as compared to gigaflops years of calculations.
\begin{table}[t]
\caption{\label{table:heavy} Ground state masses in GeV for heavy mesons.}
\begin{tabular}{||l|l@{\,}|l||l|l|l||}
 \hline\hline
 n & Experiment$^1$ & Theory    & n & Experiment$^1$& Theory    \\
 \hline\hline
 \multicolumn{3}{||c||}{$^1\mathrm{S}_0$ Singlet $\bar D^0   (u\bar c)$}  &
 \multicolumn{3}{  c||}{$^3\mathrm{S}_1$ Triplet $\bar D^{*0}(u\bar c)$} \\
 \hline
 1   & 1.8645(5) & 1.8645    & 1 & 2.0067(5)   & 1.9568    \\
     &           & 1.9224$^2$&   &             & 2.0067$^2$\\
     &           & 1.88$^3$  &   &             & 2.04$^3$  \\
 \hline\hline
 \multicolumn{3}{||c||}{$^1\mathrm{S}_0$ Singlet $B^+   (u\bar b)$} &
 \multicolumn{3}{  c||}{$^3\mathrm{S}_1$ Triplet $B^{*+}(u\bar b)$} \\
 \hline
 1   & 5.2790(5) & 5.2790    & 1 & 5.3250(6)   & 5.3082    \\
     &           & 5.2965$^2$&   &             & 5.3250$^2$\\
     &           & 5.31$^3$  &   &             & 5.37$^3$  \\
 \hline\hline
 \multicolumn{3}{||c||}{$^1\mathrm{S}_0$ Singlet $D^-_s   (s\bar c)$}  &
 \multicolumn{3}{  c||}{$^3\mathrm{S}_1$ Triplet $D^{*-}_s(s\bar c)$} \\
 \hline
 1   & 1.9685(6) & 1.9961    & 1 & 2.1124(7)   & 2.0665    \\
     &           & 2.0201$^2$&   &             & 2.0655$^2$\\
     &           & 1.98$^3$  &   &             & 2.13$^3$  \\
 \hline\hline
 \multicolumn{3}{||c||}{$^1\mathrm{S}_0$ Singlet $B^0_s   (s\bar b)$}  &
 \multicolumn{3}{  c||}{$^3\mathrm{S}_1$ Triplet $B^{*0}_s(s\bar b)$} \\
 \hline
 1   & 5.3696(24) & 5.3961    & 1 & 5.4166(35)  & 5.4185    \\
     &            & 5.3739$^2$&   &             & 5.3885$^2$ \\
     &            & 5.35$^3$  &   &             & 5.45$^3$  \\
 \hline\hline
 \multicolumn{3}{||c||}{$^1\mathrm{S}_0$ Singlet $B^+_c   (c\bar b)$} &
 \multicolumn{3}{  c||}{$^3\mathrm{S}_1$ Triplet $B^{*+}_c(c\bar b)$} \\
 \hline
 1   & 6.4(4) & 6.4914   & 1 & --- & 6.4984     \\
     &        & ---      &   &     & 6.3458$^2$ \\
     &        & 6.27$^3$ &   &     & 6.34$^3$   \\
 \hline\hline
 \multicolumn{6}{l}{ %
   $^1$Hagiwara  \textit{etal} \cite{RPP02},\ %
   $^2$Zhou and Pauli \cite{ZhoPau03b}.\ %
}\\
 \multicolumn{6}{l}{ %
   $^3$Godfrey and Isgur \cite{GodIsg85},\ %
}\\
 \multicolumn{6}{l}{ %
   $^a$To be confirmed; $^b$$J^P$ not confirmed.
}\\
\end{tabular}
\end{table}

\textbf{Strange mesons}.
The S wave $K^+$ and $K^{*+}$ spectra are given in
Table~\ref{table:us}. The mass of the ground state of $K^{+}$ is
used to determine the mass parameter $m_{s}$. 
The excitations for the $K$ (n$^1$S$_0$) differ by only 
60 and 3 MeV, respectively, 
and the spectrum for the $K^*$ (n$^3$S$_1$) by 27 and 64 MeV.
Possibly, this could even be improved by playing with the fudge parameters, 
but in view of the experimental situation, it is not done here.
Except the ground states, the experiments carry many ambiguities 
about the quantum number assignment for $K$ and $K^*$ mesons. 
Both the first and the second excited state of $K$
($2^1$S$_0$ and $3^1$S$_0$) are not confirmed. Another unconfirmed
resonance with mass $1.629\pm 0.027$ GeV lying between $2^1$S$_0$
and $3^1$S$_0$ was assigned to be a singlet $K$. Apparently there
is no position for it in the $K$ spectrum if it is an S wave
state. However, according to its mass and the present work, 
it might well be the first excited state of $K^*$ ($2^1$S$_0$).

\begin{table}[t]
\caption{\label{tab:specHeavy}
   The predicted S spectrum in GeV for heavy mesons.
}\begin{tabular}{||l|l@{\hspace{9ex}}||l|l@{\hspace{9ex}}||}
 \hline\hline
 n$^1\mathrm{S}_0$  & $\bar D^0   \qquad(u\bar c)$ &
 n$^3\mathrm{S}_1$  & $\bar D^{*0}\qquad(u\bar c)$ \\
 \hline
    1 & 1.8645 &  1  & 1.9568 \\ 
    2 & 2.5997 &  2  & 2.6666 \\ 
    3 & 3.0733 &  3  & 3.1301 \\ 
    4 & 3.4380 &  4  & 3.4889 \\ 
 \hline\hline
 n$^1\mathrm{S}_0$  & $B^+   \qquad(u\bar b)$ &
 n$^3\mathrm{S}_1$  & $B^{*+}\qquad(u\bar b)$ \\
 \hline
    1 & 5.2790 &  1  & 5.3082 \\ 
    2 & 5.9653 &  2  & 5.9911 \\ 
    3 & 6.4733 &  3  & 6.4971 \\ 
    4 & 6.8911 &  4  & 6.9134 \\ 
 \hline\hline
 n$^1\mathrm{S}_0$  & $D^-_s   \qquad(s\bar c)$ &
 n$^3\mathrm{S}_1$  & $D^{*-}_s\qquad(s\bar c)$ \\
 \hline
    1 & 1.9961 &  1  & 2.0665 \\ 
    2 & 2.6878 &  2  & 2.7405 \\ 
    3 & 3.1427 &  3  & 3.1879 \\ 
    4 & 3.4958 &  4  & 3.5365 \\ 
 \hline\hline
 n$^1\mathrm{S}_0$  & $B^0_s   \ \qquad(s\bar b)$ &
 n$^3\mathrm{S}_1$  & $B^{*0}_s\ \qquad(s\bar b)$ \\
 \hline
    1 & 5.3961 &  1  & 5.4185 \\ 
    2 & 6.0349 &  2  & 6.0549 \\ 
    3 & 6.5114 &  3  & 6.5299 \\ 
    4 & 6.9052 &  4  & 6.9227 \\ 
 \hline\hline
 n$^1\mathrm{S}_0$  & $B^+_c   \qquad(c\bar b)$ &
 n$^3\mathrm{S}_1$  & $B^{*+}_c\qquad(c\bar b)$ \\
 \hline
    1 & 6.4914 &  1  & 6.4984 \\ 
    2 & 6.9688 &  2  & 6.9752 \\ 
    3 & 7.3365 &  3  & 7.3426 \\ 
    4 & 7.6467 &  4  & 7.6526 \\ 
 \hline\hline
\end{tabular}
\end{table}

\textbf{Heavy mesons}.
The S wave $u\bar c$, $u\bar b$, $s\bar c$, $s\bar b$ and $c\bar
b$ meson spectra are given in Table~\ref{table:heavy}. No
excitations were observed for these mesons.~--- 
The mass of the ground state of $\bar D^{0}$ is used to determine
the mass parameter $m_{c}$. No much data are available for $D$ and
$D^*$ mesons. The model prediction for $1^1$S$_0$ of $\bar D^{*0}$
is smaller than experiment by 50 MeV.~---
The mass of the ground state of $\bar B^{+}$ is used to determine
the mass parameter $m_{b}$. The ground state of $\bar B^{*+}$ 
agrees with the present model to within 17 MeV.~---
No experimental values in the $s\bar c$ mesons are used to
determine the model parameters. The model deviates from the
available ground states by 27 and 45 MeV.~---
No data in $s\bar b$ mesons are used to determine the model parameters. 
The model deviates from the experiment by 26 and 2 MeV, 
and is thus almost within the experimental errors.~---
The mass of the ground state of $B^+_c$ ($1^1$S$_0$) carries a
large experimental error. Model and experiment agree.~---
The model prediction for the excited states are compiled
in Table~\ref{tab:specHeavy}, for easy reference.

\section{Conclusions}

The agreement between the present simple model
and the experiment is excellent, with small but significant deviations. 
Perfect agreement has not been the goal of the present work. There must 
be room for a possible improvements by the `true' equation (\ref{eq:5}).

With the 4 mass parameters of the up/down, strange, charm and
bottom quarks, the model has only 2 additional parameters 
for the linear potential. In principle, the fudge
factors should be counted as parameter as well, 
but as seen above, the choice of the up/down mass and 
the fudge factors is strongly coupled.
Thus, with 6 canonical parameters  the model exposes a reasonably good
agreement with all 21 available data points.

Note that renormalized gauge field theory has also 4+1+1
parameters:
The 4 flavor quark masses, the strong coupling constant $\alpha_s$,
and the renormalization scale $\lambda$.
Of course, they can be mapped into each other \cite{Pau03a,Pau03b}.

Once one has determined the parameters in such a first guess, 
one should relax the model assumption, Eq.(\ref{ieq:8}),
and work with the full non local model, with a position
dependent mass. For this one has to go back to the computer
and perform the necessary fine tunings of the parameters. 

\clearpage
\begin{appendix}
\section{Solution to the linear potential}
\label{asec:1}
Restricting to spherical symmetry ($l=0$), 
the Schr\"odinger equation for $V(r)=Fr$ is
\begin{eqnarray}
   \begin{array}{rcl}
   \Big[\frac{\vec p^2}{2m_r}+Fr-E_n\Big] \Psi_n(r) &=& 0 
\,,\\
   \Big[-\frac{1}{2m_r}\frac{1}{r}
   \frac{d^2}{d r^2} r+Fr-E_n\Big]\Psi_n(r) &=& 0
\,,\\
   v_n''(r)-2m_r(Fr-E_n)v_n(r) &=& 0
\,,\end{array}
\label{aeq:8}\end{eqnarray}
In the last step  
$v_n(r) =  r\Psi_n(r)$ was substituted. 
With the dimensionless variable
\begin{eqnarray}
   \xi = [2m_rF]^{\frac13} r - \xi_n
\,,\end{eqnarray}
equation (\ref{aeq:8}) is mapped into 
the differential equation for the 
Airy function $u_n''(\xi)-\xi u_n(\xi)=0$.
Eigenvalues are obtained by boundary
conditions, in this case $v_n(0)=0$, 
\begin{eqnarray}
   \begin{array}{rcl}
   E_n &=& \bigg[\frac{F^2}{2m_r}\bigg]^{\frac13}\,\xi_n \equiv 
   \hbar\omega\,\xi_n 
\,,\\
   \Psi_n(r) &=& \frac1r \mathrm{Ai}\big(r[2m_rF]^{\frac13} - \xi_n\big)
\,.\end{array}
\label{aeq:10}\end{eqnarray}
The $\xi_n$ are the negative zeros of the Airy functions \cite{HMF}: 
\begin{eqnarray}
   \begin{array} {r@{\qquad}c@{\qquad}c@{\qquad}c}
      n & \xi_n   & \eta_n  & \delta_n \\
      0 & 2.33811 & 0       & 0        \\
      1 & 4.08795 & 1.74984 & 1.74984\\
      2 & 5.52056 & 3.18245 & 1.43261 \\
      3 & 6.78670 & 4.44853 & 1.26614 
   \end{array}
\label{aeq:11}\end{eqnarray}
The table includes $\eta_n = \xi_n-\xi_0$. 
The table includes also the spacings $\delta_n = \xi_n-\xi_{n-1}$, 
which vary slowly with $n$.
\end{appendix}
\end{document}